# Canonical Turbulence Theory

T.-W. Lee

Mechanical and Aerospace Engineering, SEMTE

Arizona State University, Tempe, AZ 85287

**Abstract**

A theoretical analysis is presented for turbulent flows, applicable for canonical (channel, boundary-layer and free jet) geometries. Momentum and energy balance for a control volume moving at the local mean velocity decouples the fluctuation from the mean velocities, resulting in a symmetric set of transport equations for the Reynolds normal and shear stresses. In this formalism, gradients of the fluctuating velocities represent flux vectors, easily verifiable using the available DNS data. A derivative of this transport concept is the scaling for the Reynolds stresses in the "dissipation" space. Combining with the statistical energy distribution function, a full prescription of turbulent flows is enabled in the basic canonical geometries. Based on this theoretical foundation, more complex flow configurations may be addressed with far more efficient algorithms.

# INTRODUCTION

Serious attempts at theoretically solving the turbulence problem have been intermittent from the era of the Bousinesque hypothesis [1], which served a practical purpose but altered the trajectory of the genre in science and engineering. Rapid advances in numerical and experimental fluid dynamics have produced vast amounts of data [e.g. 2-8], with which some of the underlying physics can be observed, and new theories be tested. A recent formalism provides radically different but succinct perspectives on the internal dynamics [9-12] and spectral structure of turbulence [13, 14]. There are attendant fundamental and practical implications, e.g. computability of important phenomena such as the Earth climate dynamics, and determination of the complete energy distributions in turbulent flows.

Turbulence in canonical flow geometries can be viewed from an alternative coordinate frame, moving at the local mean velocity. Intuitively and algebraically, this would lead to the following symmetrical transport equations for the Reynolds stress components, $u'^2$, $v'^2$, and $u'v'$. The full derivation and efficacy in reproducibly computing the turbulence structure for wall-bounded and free jet flows have been demonstrated in prior works [9-12]. The premises of the theory are that the momentum and energy are conserved in non-inertial coordinate frame moving at the local mean velocities, and that the fluctuation variables represent flux terms. For example, $du'^2/dy$ and $d(u'v')/dy$ in Eq. 1 are be viewed as the streamwise and lateral (cross-stream) transport of $u'$ momentum, respectively, balanced by the pressure ($C_{12}$-term) and viscous ($C_{13}$) forces. In their simple balanced forms, the inter-dynamics of the variables are prescribed in triadic coupled equations:

u' momentum transport:

$$\frac{d(u'v')}{dy} = -C_{11} U \frac{d(u'^2)}{dy} + C_{12} U \frac{dv'^2}{dy} + C_{13} \frac{d^2 u'}{dy^2} \tag{1}$$

v' momentum transport:

$$\frac{d(v'^2)}{dy} = -C_{21} U \frac{d(u'v')}{dy} + C_{22} U \frac{dv'^2}{dy} + C_{23} \frac{d^2 v_{rms}}{dy^2} \tag{2a}$$

u'² kinetic energy transport:

$$\frac{d(u'^3)}{dy} = -C_{31} \frac{1}{U} \frac{d(u'v' \cdot u')}{dy} + C_{32} \frac{1}{U} \frac{d(v' \cdot u'v')}{dy} + C_{33} \frac{1}{U} \left(\frac{du'}{dy}\right)^2 \tag{3}$$

In Eqs. 1-3, the fluctuating terms, u'v', v'², etc., are implicitly Reynolds-averaged. From the perspective afforded by this formalism, a realization comes about that it makes far more fluid-dynamical sense to relate the momentum and energy interchanges occurring among fluctuating variables, as opposed to coercing a relationship with mean quantities (e.g. u'v' ~ dU/dy) or dealing with a large number of unknown triple correlations in Eulerian Reynolds-stress models [13]. The concepts are anti-thematic to the classical turbulence literature to date, but the above transport dynamics provides succinct yet full insights on the turbulent transport physics, in addition to providing direct quantitative prescriptions of the internal transport processes [9-12].

One of the derivatives of this approach is the discovery of unified scaling of the Reynolds stress gradients [12], promulgated from the precise inter-relationship between the *gradient* structures. As we shall see shortly, each of the turbulence variables, $u'^2$, $v'^2$ and $u'v'$, is collapsible into a single profile in the "dissipation space". This has a profound implication in that only a unitary solution to the above set of equations is required. Multiplicative scaling operations transpose the profiles to other Reynolds numbers and in some instances across flows of similar geometries [12], with some requisite modifications for the far-field structure and outer boundary conditions. In this work, we would like to integrate these concepts onto a unified turbulence theory, applicable for canonical (channel, boundary-layer, and jet) flows. This will serve as a foundation for integration into a computational protocol for flows in more complex geometries.

**DISSIPATION SCALING**

We have shown in recent works that gradient structures in wall-bounded flows are self-similar [12]. After normalizing by local extrema, $du'^2/dy+$, $d^2v'^2/dy+^2$, and $d^2u'v'/dy+^2$, collapse onto single respective curves, as shown in Figs. 1-3. Similar to the abbreviated notations in Eq. 1-3, here $u'^2$ represents $<u+^2>/u_\tau^2$, etc; fluctuation velocities are Reynolds-averaged and normalized by the friction velocity squared. The secondary peaks (-/+) have also been ratioed relative to the first (+/-), so that the same peak-to-peak amplitudes are retained in the profiles. There are some deviatoric data points near the wall, possibly due to insufficient numerical resolution, and also in the far field caused by the subtle differences in the outer boundary conditions. Nonetheless, the unification of the profiles is nearly complete, at least at these Reynolds numbers, and applies across wall-bounded geometries: channel (CF) and boundary-layer flow over a flat plate (FP). The scaling factors simply are functions of the Reynolds number, but asymmetrical, as shown in Fig. 4. It is interesting to note that $u'^2$-structure scales at the first-gradient level, while $v'^2$ and $u'v'$ do so at the second. The fact that Eqs. 1 and 2 are momentum-conserving, while Eq. 3 is an expression of the energy balance, may have some bearing on this: momentum diffuses proportionately to the second-gradient, whereas energy dissipation has a linear dependence on the first-derivative squared. This is subject to further contemplations, starting from the next section. For the free jet flows, r/x or y/x scaling already accomplishes the same effect [14], so that a single solution set suffices [11].

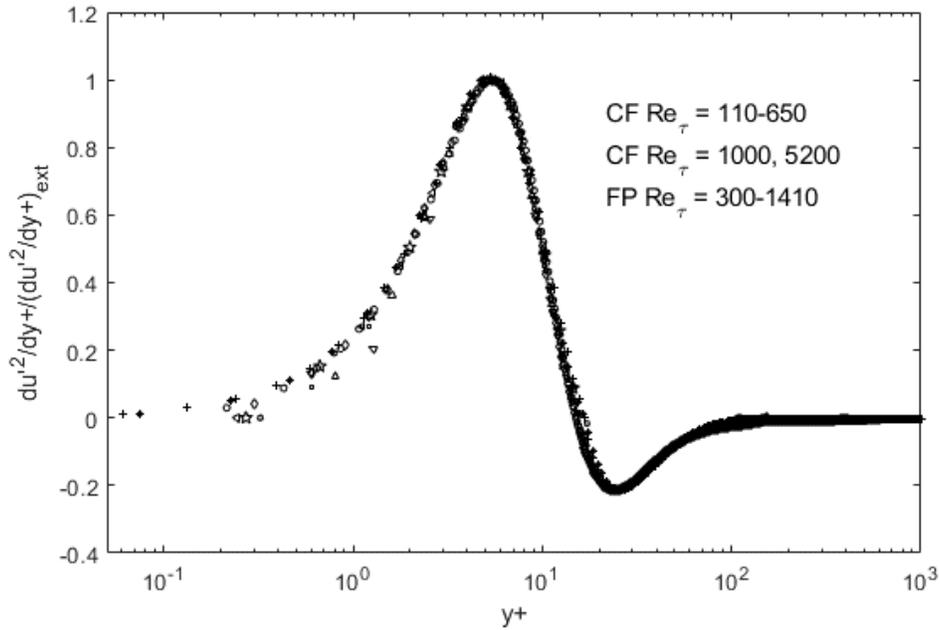

**Figure 1. Scaling of the u′² structure, at the first-gradient level. DNS data from Refs. 4 (CF) , 5 (CF) and 8 (FP) are used. CF=channel flow; FP=zero-pressure gradient boundary layer flows over a flat plate. The peaks are normalized by the local extrema, then ratioed by their relative magnitudes.**

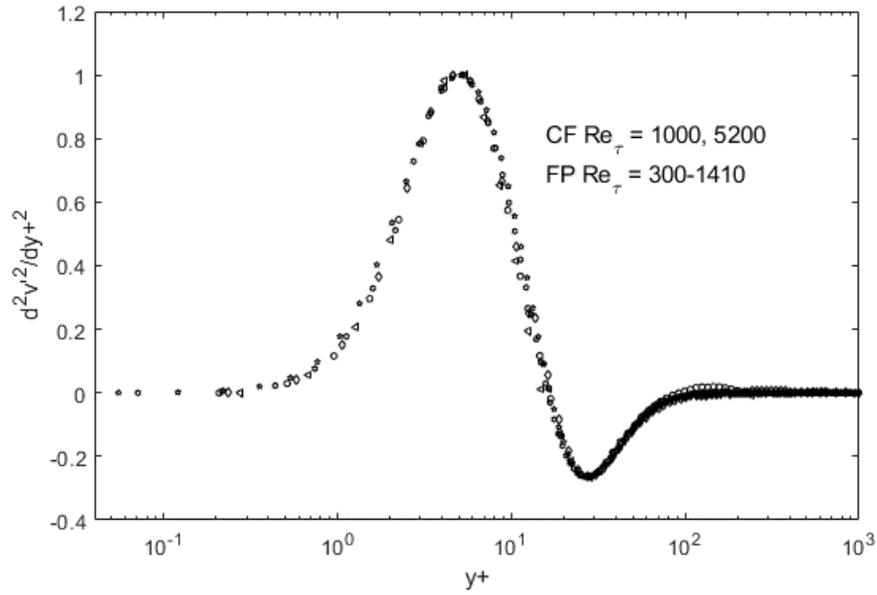

**Figure 2. Scaling of the $v'^2$ structure, for the second gradients. DNS data from Refs. 5 (CF) and 8 (FP) are used. The peaks are normalized by local extrema, then ratioed by their relative magnitudes.**

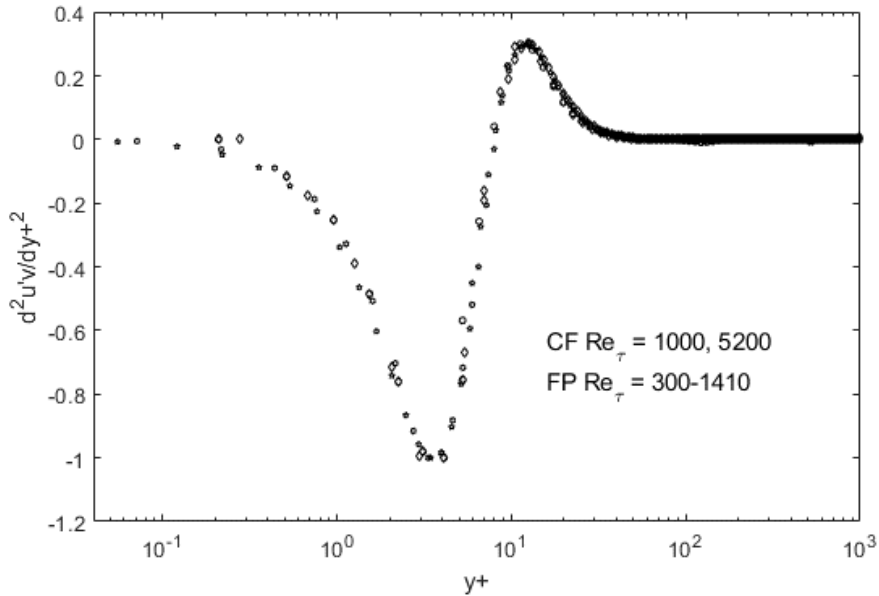

**Figure 3. Scaling of the $u'v'$ structure, for the second gradients. DNS data from Refs. 5 (CF) and 8 (FP) are used. The peaks are normalized by the local extrema, then ratioed by their relative magnitudes.**

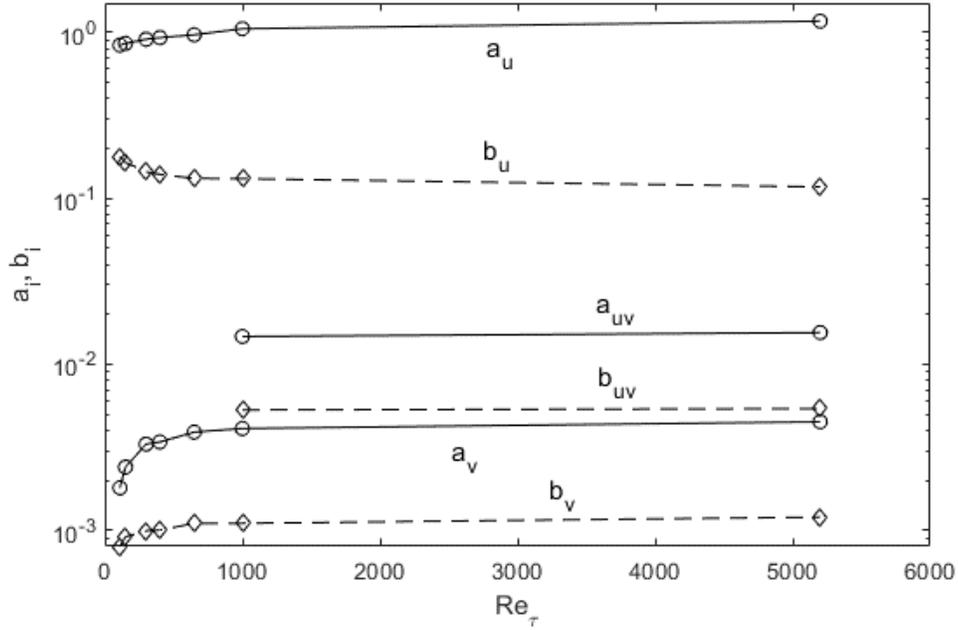

**Figure 4. Scaling factors for the above profiles. ($a_u$, $b_u$)= magnitude of the (peak, nadir) for the $du'^2/dy+$ structure; ($a_v$, $b_v$)= magnitude of the (peak, nadir) for the $d^2v'^2/dy+^2$; and ($a_{uv}$, $b_{uv}$)= magnitude of the (nadir, peak) for the $d^2u'v'/dy+^2$ structure.**

In addition to the asymmetrical order in the dissipation structure, we can see that shape of the gradient profiles appears to be translatable, with some lateral shift, from one (e.g. $du'^2/dy+$) to the other ($d^2v'^2/dy+^2$). To illustrate, the same gradients plotted in Figs. 1-3 can be collected, for one Reynolds number, and plotted in a single graph (Fig. 4a), wherein $d^2v'^2/dy+^2$ resembles $d^2u'v'/dy+^2$, except pushed further away from the wall. $d^2u'v'/dy+^2$ is a sort of mirror reflection of $du'^2/dy+$, this time compressed toward the wall. An alternate collection of all the first gradients, $du'^2/dy+$, $dv'^2/dy+$, and $du'v'/dy+$, in Fig. 4b, is also interesting. As noted above, both physically, and now visually, there is likely to be relationships among the fluctuation variables, at the first- or second-gradient levels.

Upon these observations, modelers may have attempted u'v'-approximations based on a function series of the fluctuation gradients. From a theoretical incline, however, a more direct path is to start from the conservation principles in a non-inertial coordinate frame, as we have done [9-12], which would bring us to the terminus in the form of Eqs. 1-3. Thus, the self-similarities exhibited in Fig. 4 are both hints and manifestations of the inter-dynamics or transport processes among the root turbulence variables, as visualizable in Figs. 1-3 and encapsulated in Eqs. 1-3.

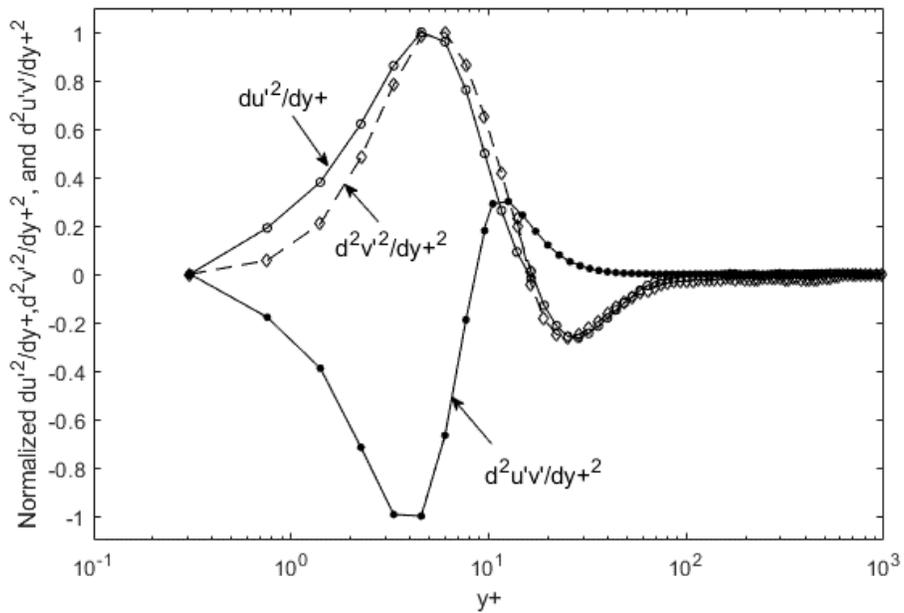

(a)

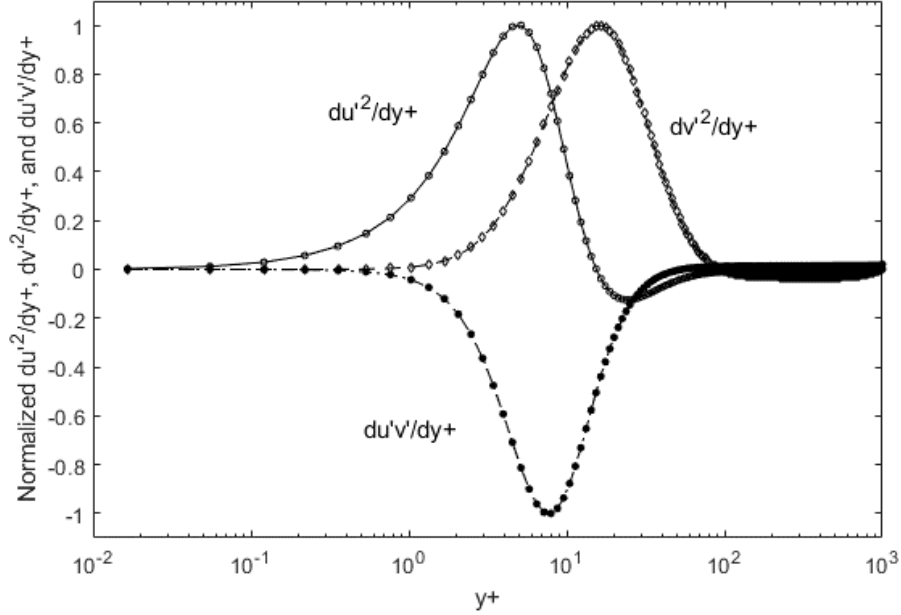

(b)

**Figure 5. Comparisons of the similarities in the gradient profiles for $u'^2$, $v'^2$, and $u'v'$. (a) Mixed first- and second-gradients; and (b) all first gradients. DNS data are from Graham et al. [5] for $Re_\tau = 1000$.**

**TURBULENCE FLUX DYNAMICS**

We should examine the inter-transport dynamics in some details. Since the Reynolds shear stress equation (Eq. 1) has been amply validated in previous work [9-12], to avoid overlap, let us focus on $du'^3/dy+$ (Eq. 3) and $dv'^2/dy+$ (Eq. 2). Both of these fluxes are important contributors to the Reynolds shear stress in Eq. 1, and scripting these two root variables would suffice in full reconstruction of $u'v'$, de-necessitating any turbulence models. Eq. 3 shows that $u'v'$ and $v'^2$ in turn combine to prescribe the kinetic energy flux gradient, $du'^3/dy$, forming a triad of energy terms to produce the observed turbulence structure. Here, we take the Lagrangian interpretation of $u'^3$ as $(u'^2)^{3/2}$, just as $u'_{rms}$ is set equal to $(u'^2)^{1/2}$ in Eq. 1, so that $u'^2$ profiles can be obtained directly from integration of $du'^3/dy+$. We can see in Fig. 6 that the agreement between the current theory (Eq. 3) and DNS data [5] is quite good, until we reach the "far-field", $y+ > 30$. The main composition of the kinetic energy flux are the lateral transport ($C_{31}$ term) and the pressure work ($C_{32}$). The viscous dissipation ($C_{33}$) term is quite small at this Reynolds number ($Re_\tau = 1000$), and its omission does not subtract from the current accuracy. Thus, a succinct visualization emerges wherein the streamwise flux of kinetic energy, $du'^3/dy+$, is balanced by the cross-stream transport ($C_{31}$) and expended through pressure work ($C_{32}$) and viscous dissipation ($\ll 1$). It is a simple matter to numerically integrate and see that this gradient structure will result in a sharp peak near the wall, followed by an increasingly abrupt bend toward gradual decline to the centerline boundary condition [12].

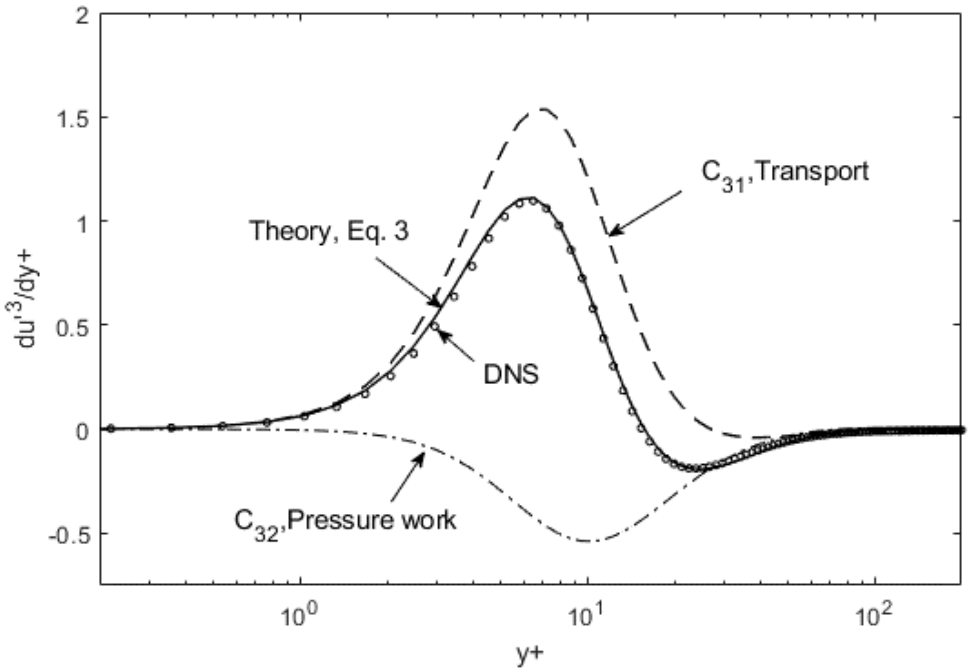

**Figure 6. Energy balance of Eq. 3, leading the turbulence kinetic energy structure. DNS data are from Graham et al. [5] for $Re_\tau = 1000$.**

Let us examine what is going in the far field, using zoomed-in visualization in Fig. 7. At large Reynolds numbers, secondary peaks in $u'^2$ start to appear [15], and although we do not have access to high-resolution DNS data at this time we can observe the beginning stage of such secondary structures. Upon close examination, we can see that a weak undulating pattern develops near y+ ~ 100 for $Re_\tau$=5200, in contrast to a monotonic approach to zero outer boundary condition at $Re_\tau$=1000. This was attributed to requisite increase in the dissipation at high Reynolds numbers [12], similar to a second harmonic generation of wave energy. We can project that as the "second harmonic" becomes amplified (and/or shifted upward) the undulation may protrude to positive in Fig. 7, thus causing the observed secondary peaks. It is evident that this would not be subject to the same simple scaling applicable prior to this point (left of the vertical dash line in Fig. 7), so that an alternate secondary scaling appears to be necessary to capture this effect in the far field. The theoretical lines are for the $Re_\tau$=5200, and it illustrates the identical dynamic balance of the transport and pressure work in the energy equation (Eq. 3), leading to the initiation of the secondary structure. In this region (y+>100), the transport ($C_{31}$ term) is negative, whereas the pressure work turns slightly positive. It is extrapolatable that at higher Reynolds numbers the $C_{32}$ effect would bring the overall slope ($du'^3/dy+$) briefly to a minute plus side, resulting in the secondary peaks observed in experiment and DNS [15].

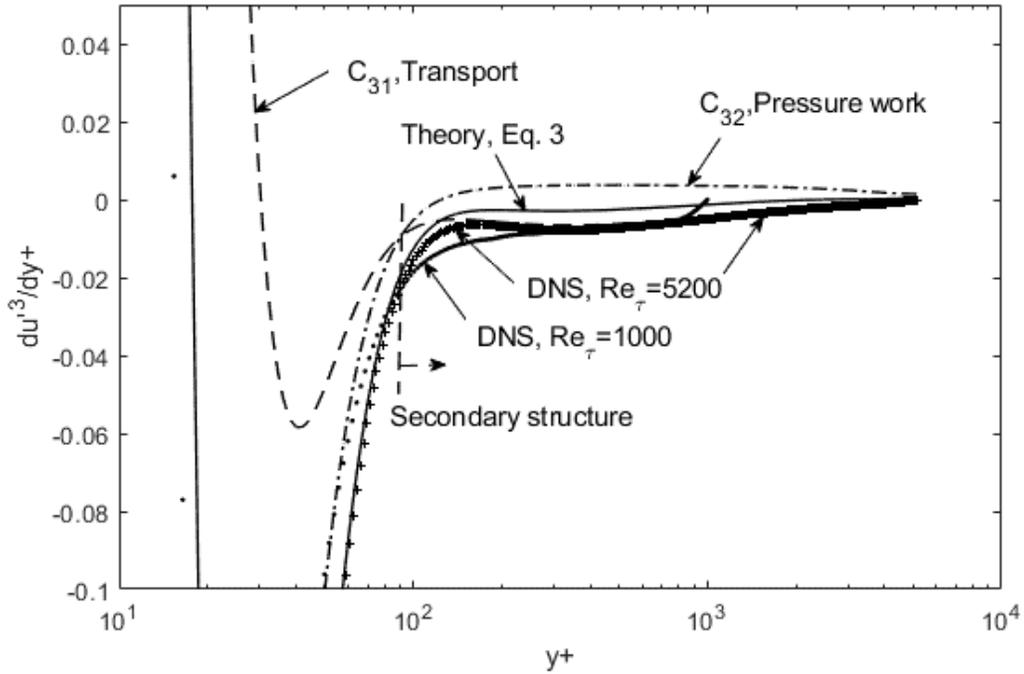

**Figure 7. Far-field characteristics in the turbulence kinetic energy profile, showing the nascent secondary structure. DNS data are from Graham et al. [5].**

Returning to the global structure for $dv'^2/dy$ in Fig. 8, again the set of flux terms of Eq. 2 prescribe the composition of the v'-momentum content accurately, thus completing the specifications of the root turbulence variables in canonical turbulent flows. As noted above, with the $u'^2$ and $v'^2$-gradients in Eq. 1, the Reynolds shear stress, u'v' can be directly or iteratively computed, which upon substitution in the Reynolds-averaged Navier-Stokes equation results in the mean velocity profile [11, 12]. For jet flows, the situation is made yet simpler due to absence of the pressure terms in Eqs. 1-3, and solutions scaled by y/x are iteratively achieved [11].

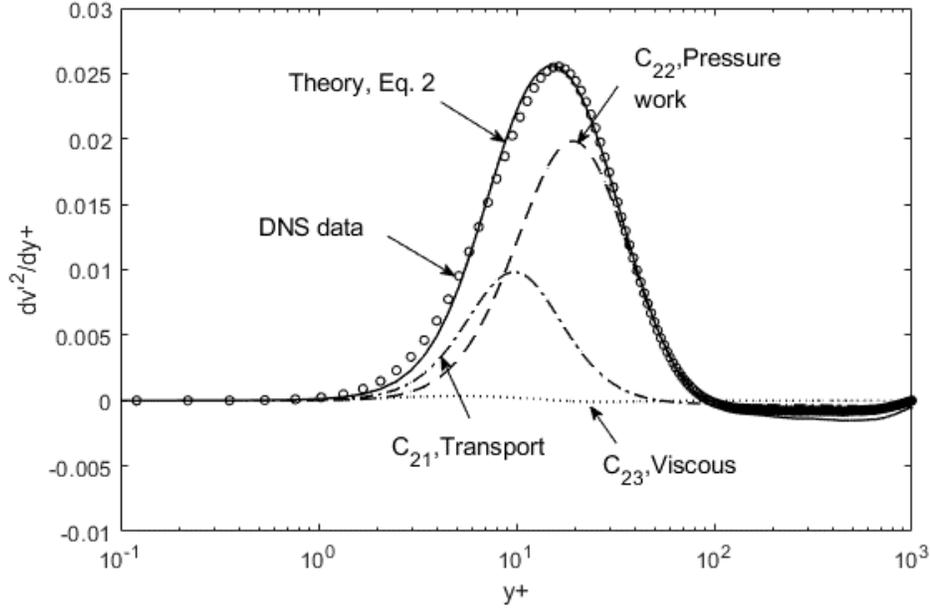

**Figure 8. Momentum budget of Eq. 2, leading the net lateral momentum flux, dv′²/dy+. DNS data are from Graham et al. [5] for Re$_\tau$= 1000.**

Finally, a closely related statistical component is the energy distribution or spectra. Since u′² and v′² represent the mean turbulence kinetic energy, enumeration of its spectral content will complete the statistical prescription. In this regard, we drew upon yet another fundamental principle of the Second Law, in the corollary form of the maximum entropy method [16, 17], which derived the full spectral function, parameterizable with the root turbulence variables, as shown below.

$$E(k) = \frac{C_1}{k^4} exp\{-C_2 u'^2 - C_3 k^2 u'^2\} \qquad (4)$$

This spectral function, along with the kinematic scaling for u' ~ m-log(k), is found to be applicable in two-, three-dimensional homogeneous turbulence, and in channel flows [16, 17], reproducing the full observed energy distributions continuously from the energy-containing, inertial to the viscous dissipation range. Further confirmation would be useful in other turbulence geometries, but the above set of findings constitute a foundation for theoretical analysis of canonical turbulence.

**CONCLUDING REMARKS**

A formalism is summarized in which turbulence transport can be dynamically described in canonical flow geometries through a set of symmetry equations, derivable from the momentum and energy balance for a control volume moving at the local mean velocity. Gradients of the fluctuating velocities represent flux vectors, balanced by force or energy terms, as verifiable using the available DNS data. The insights gained from this perspective lead to a unified scaling for the Reynolds stresses in the "dissipation" space. Combining with the statistical energy distribution function, a full specification of turbulent flows is enabled in the basic canonical geometries. This approach is evidently distinguishable from post-analytical methods such as attached-eddy hypothesis [18-20], and results in the complete structural descriptions based on physical principles. Also, some scalable trends are observable on the facet [15], but unified structures are found below the surface at the first- and second-gradient levels. Philosophical departure from the confines of orthodox lineage, in this fortuitous instance, leads to a disruptive, but definitive formalism on turbulence. Based on this theoretical foundation, utilizations in more complex flow configurations can also be envisioned.

# REFERENCES

[1] Boussinesq, T. V., 1877, *Mém. pres. Acad. Sci.*, 3rd ed., Paris XXIII, p. 46.

[2] Moin, P. and Mahesh, K., 1998, Direct numerical simulation: A tool in turbulence research, Annual Review of Fluid Mechanics, Vol. 30, pp. 539-578.

[3] Mansour, N.N., Kim, J. and Moin, P., 1998, Reynolds-stress and dissipation-rate budgets in a turbulent channel flow, Journal of Fluid Mechanics, Vol. 194, pp. 15-44.

[4] Iwamoto, K., Sasaki, Y., Nobuhide K., 2002, Reynolds number effects on wall turbulence: toward effective feedback control, International Journal of Heat and Fluid Flows, 23, 678-689.

[5] Graham, J., Kanov, K, Yang, X.I.A., Lee, M.K., Malaya, N, Lalescu, C.C., Burns, R., Eyink, G, Szalay, A, Moser, R.D., and Meneveau, C., 2016, A Web Services-accessible database of turbulent channel flow and its use for testing a new integral wall model for LES, Journal of Turbulence, 17(2), 181-215.

[6] Na, Y. Hanratty, T and Liu, T.J., 2001, The use of DNS to define stress producing events for turbulent flow over a smooth wall, Flow, Turbulence and Combustion, 66, pp. 495–512

[7] Moser, Robert D., John Kim, and Nagi N. Mansour, 1999, Direct numerical simulation of turbulent channel flow up to $Re_t$= 590, Physics of Fluids, 1999, 11,4, pp. 943-945.

[8] Spalart, P.R., 1988, Direct simulation of a turbulent boundary layer up to Re=1410, Journal of Fluid Mechanics, Vol. 187, pp. 61-77.

[9] Lee, T.-W., 2018, Reynolds stress in turbulent flows from a Lagrangian perspective, Journal of Physics Communications, 2, 055027.

[10] Lee, T.-W., and Park, J.E., Integral Formula for Determination of the Reynolds Stress in Canonical Flow Geometries, Progress in Turbulence VII (Eds.: Orlu, R, Talamelli, A, Oberlack, M, and Peinke, J.), pp. 147-152, 2017.

[11] Lee, T.-W., 2020, Lagrangian Transport Equations and an Iterative Solution Method for Turbulent Jet Flows, Physica D, 132333.

[12] Lee, T.-W., 2021, Asymmetrical Order in Wall-Bounded Turbulent Flows, submitted to a physics journal. An alternate version is viewable and citable as, arxiv:2102.00488.

[13] Hanjalic, K., 1994, Advanced turbulence closure models: a view of current status and future prospects, Int. Journal of Heat and Fluid Flow, Vol. 15, No. 3, 1994, pp. 178-203.

[14] Hinze, J.O., *Turbulence*, McGraw-Hill Series in Mechanical Engineering. McGraw-Hill, New York, 1975.

[15] Marusic, I., McKeon, B.J., Monkewitz, P.A., Nagib, H.M., Smits, A.J., and Sreenivasan, K.R., 2010, Wall-bounded turbulent flows at high Reynolds numbers: Recent advances and key issues, Physics of Fluids, 22, 065103.

[16] Lee, T.-W., 2020, Lognormality in turbulence energy spectra, Entropy, 22(6), 669.

[17] Lee, T.-W., 2021, Scaling of maximum entropy turbulence energy distribution, European Journal of Mechanics, B/Fluids, 87, pp. 128-134.

[18] R. J. Adrian, "Closing in on models of wall turbulence," Science **329**, 155–156 (2010).

[19] Han, J., Hwang, J., Yoon, M., Ahn, J. and Sung, H. J., 2019, Azimuthal organization of large-scale motions in a turbulent minimal pipe flow, Physics of Fluids 31, 055113.

[20] Marusic, I. and Monty, J.P., 2019, Attached eddy model of wall turbulence, Annual Review of Fluid Mechanics, 51:49-74.